\documentclass{amsart}

\usepackage{amssymb,latexsym,amsfonts,amsmath}
\usepackage{multicol} 
\usepackage{graphicx}
\include{diagrams}
\usepackage{eso-pic}
\usepackage{epsfig}
\usepackage{graphicx}
\usepackage{color}
\usepackage{tikz}
\usetikzlibrary{automata}
\usetikzlibrary{shapes}
\usepackage[normalem]{ulem}
\usepackage{refcount}

\usepackage{amssymb,latexsym,amsfonts,amsmath}
\usepackage{graphicx}

\newtheorem{theorem}{Theorem}[section]

\newtheorem{proposition}[theorem]{Proposition}

\newtheorem{definition}[theorem]{Definition}

\numberwithin{equation}{section}

\def\myparagraph#1{{\smallskip\noindent{\bf #1}}}

\newcommand{\R}{{\mathbb{R}}}

\newcommand{\N}{{\mathbb{N}}}

\def\set#1{{ \{ #1 \}}}
\def\tuple#1{{ \langle #1 \rangle}}
\def\true{{\mathit{true}}}

\newcommand{\refine}[1]{\Delta}

\def\nats{\mathbb{N}}

\begin{document}

\begin{abstract}
Software implementations of controllers for physical systems are at the core of many embedded systems.
The design of controllers uses the theory of dynamical systems to construct a 
mathematical control law that ensures that the controlled system has certain properties, 
such as asymptotic convergence to an equilibrium point, while optimizing some performance criteria.
However, owing to quantization errors arising from the use of fixed-point arithmetic,
the implementation of this control law can only guarantee
{\em practical} stability: under the actions of the implementation, the trajectories
of the controlled system converge to a bounded set around the equilibrium point, and the
size of the bounded set is proportional to the error in the implementation.
The problem of verifying whether a controller implementation achieves practical stability
for a given bounded set has been studied before.
In this paper, we change the emphasis from verification to automatic {\em synthesis}. 
Using synthesis, the need for formal verification can be considerably reduced thereby reducing 
the design time as well as design cost of embedded control software. 

We give a methodology and a tool to synthesize embedded control 
software that is Pareto optimal
w.r.t.\ both performance criteria and practical stability regions.
Our technique is a combination of static analysis to estimate quantization 
errors for specific controller implementations
and stochastic local search over the space of possible controllers using particle swarm optimization.
The effectiveness of our technique is illustrated using examples of
various standard control systems: 
in most examples, we achieve controllers with close
LQR-LQG performance but with implementation errors, hence regions of practical stability, 
several times as small.
\end{abstract}

\title[Synthesis of Minimal Error Control Software]{Synthesis of Minimal Error Control Software}

\author[Rupak Majumdar]{Rupak Majumdar$^1$} 
\author[Indranil Saha]{Indranil Saha$^2$} 
\author[Majid Zamani]{Majid Zamani$^3$}
\address{$^1$Max Planck Institute for Software Systems\\
Kaiserslautern, Germany}
\email{rupak@mpi-sws.org}
\urladdr{http://www.cs.ucla.edu/~rupak}
\address{$^2$Department of Computer Science\\
University of California at Los Angeles,
Los Angeles, CA 90095}
\email{indranil@cs.ucla.edu}
\urladdr{http://www.cs.ucla.edu/~indranil}
\address{$^3$Department of Electrical Engineering\\
University of California at Los Angeles,
Los Angeles, CA 90095}
\email{zamani@ee.ucla.edu}
\urladdr{http://www.ee.ucla.edu/~zamani}

\maketitle

\section{Introduction}

Software implementations of controllers for physical systems are the core of
many critical cyber-physical systems.
The design of these systems usually proceeds in two steps.
First, starting with a mathematical model of the system, one designs a
mathematical control law 
that ensures that the physical system, equipped with this control law, has certain
desirable properties such as asymptotic stability (convergence to an ideal behavior)
and performance.
Second, the control law is implemented as a software task on a specific hardware architecture.
Since the implementation has quantization errors due to the use of 
fixed-precision representation of real numbers, the quantization of a stabilizing controller may 
lead to limit cycles and chaotic behavior \cite{kalman}. Hence, the implemented system usually 
guarantees the weaker property of {\em practical} stability,
where the system is guaranteed to converge to a bounded set around the ideal behavior
and the size of the bounded set is proportional to the quantization error. 

Much recent research has focused on verifying that
a given implementation of a control law
guarantees that the practical stability region lies within a given set
\cite{PodelskiW06,PodelskiW07,feron_journal,anta,DM11}.
In this paper, we change the emphasis from verification to {\em synthesis}.
We provide a design methodology to synthesize a control implementation for which 
the effect of implementation errors on system performance is minimized. 
Hence, the need for verification can be substantially reduced.

We focus on linear systems in this paper.
For linear systems, a standard optimal control design approach uses 
the {\em linear quadratic regulator} (LQR) and {\em linear quadratic Gaussian} (LQG) algorithms~\cite{joao}, 
which finds a feedback controller stabilizing the plant while minimizing quadratic cost functions. 
The LQR cost function takes into account the deviations of the state and control inputs from ideal
values and the LQG cost function takes into account the deviation of the state from its estimation. 
However, in general, they do not take implementation errors arising from
fixed-precision arithmetic into account.
Thus, a controller optimizing only the LQR-LQG cost may have a large implementation error
because its implementation on a fixed-precision platform has large numerical
errors, but a controller ``close'' to the optimal performance may have much lower
numerical errors when implemented on the same platform.

In our methodology, we modify the LQR-LQG performance criterion to additionally minimize 
the error due to quantization in the implementation.
Technically, our methodology has two parts.
First, how can we estimate the quantization error of a given implementation?
Second, how can we find Pareto-optimal points for the two objectives given by the 
LQR-LQG and quantization error cost functions?
We proceed as follows.

For the first step, 
for a given linear feedback controller and the operating intervals of the states of the plant and the controller, 
we first perform a precise range analysis of the controller variables, 
and use the computed ranges to allocate bitwidths to each controller variable. 
We implement our range analysis based on linear programming. 
Using the allocated bitwidths, we generate code for a fixed-precision 
program implementing the control law. 
Finally, we use an algorithm based on mixed-integer linear programming to find 
a bound on the maximum difference 
between the ideal control law and the output of the fixed-precision program.

For the second step, we optimize a weighted linear combination of the two cost functions
using a stochastic local search technique.
LQR-LQG is attractive because it gives rise to a {\em convex} optimization problem,
for which efficient solutions are known.
Unfortunately, additionally tracking the quantization error results in a non-convex
optimization problem.
We solve the non-convex optimization problem using {\em particle swarm optimization} (PSO), a
population-based stochastic optimization approach \cite{kennedy,liu,jiang}. 
PSO iteratively solves an optimization problem by maintaining a population (or {\em swarm}) of candidate controllers, 
called {\em particles}, and moving them around in the search-space of possible
controllers, trying to minimize the objective function.

In more detail, our algorithm proceeds as follows.
Given a linear control design problem, we set up a non-convex optimization problem to minimize a weighted combination
of the LQR-LQG cost function and the implementation error.
We minimize this cost function using PSO.
In each step of PSO, given a new position of a particle, we check if the position represents a stabilizing
controller (by examining the eigenvalues of the controlled system).
If not, we assign the position an infinite cost.
If the position represents a stabilizing controller, we generate the best possible 
fixed-point code for this controller under a hardware budget
and perform static analysis to estimate a bound on the implementation error.
We compute the value of the objective function by taking the weighted sum of the LQR-LQG cost and this bound.
We continue PSO until convergence or until some iteration bound is met.
At this point, we output the controller that minimized the objective function.

We have implemented this methodology
on top of Matlab's Control Theory Toolbox, using an implementation of PSO proposed in~\cite{ebbesen}, 
and a custom static analysis using the lp\_solve linear programming
tool. 
In our experiments, we compare the LQR-LQG cost and implementation errors of
controllers generated by conventional LQR-LQG optimization (implemented in Matlab)
with controllers generated by PSO using our methodology.
In most cases, our controllers have LQR-LQG costs close to the optimal LQR-LQG controllers, but
have implementation errors that are reduced by a factor of 4 or more.
Thus, we generate controllers with guaranteed bounds on practical stability regions that are
4 times or more smaller than the pure LQR-LQG controllers.
Our work provides, for the first time, an integrated analysis and tool to take 
quantization errors into account in model-based design and implementation of controllers.

\myparagraph{Other Related Work}
Besides the related work mentioned above, we mention
the results in \cite{williamson,williamson1,liu1} which provide controller synthesis approaches minimizing some performance 
criteria while controllers are implemented using fixed-point arithmetic. 
The results in \cite{williamson,williamson1,liu1} assume some excitation conditions under which the quantization error 
can be modeled as a zero mean uniform white noise. 
Furthermore, they do not provide any bounds on regions of practical stability. 
However, the result in this paper does not have any assumption on the quantization error 
and it provides an explicit bound on the regions of practical stability.

The range analysis problem has been studied extensively in the context of 
optimum bitwidth allocation to intermediate variables in a fixed-point program, mostly in the DSP domain.
Both static \cite{LGCMLC_TCAD_06, LCN_TCAD_07, OCCLM_FPL_07} 
and simulation-based \cite{BR_RSP_05, MSBZ_TCAD_07} approaches have been used. 
Static approaches usually employ abstractions based on interval arithmetic \cite{Moore66} or affine arithmetic \cite{SF_IMPA_97}. 
Simulation-based methods, especially those performing constrained-random simulations, suffer from the lack of completeness. 
This causes the resultant system to be non-robust, incomplete simulation can lead to overflow 
conditions resulting in incorrect behavior.
We found our mixed-integer linear programming approach to be both precise and reasonably fast for our application.

\section{Preliminaries}\label{preliminaries}

\subsection{Controllers and Observers}
We use symbols $\N_0$, $\mathbb{R}$, and $\mathbb{R}^+_0$ for the set of nonnegative integers, real and nonnegative real numbers.
For a vector $x\in\mathbb{R}^{n}$, we denote by $x_{i}$ the {$i$-th} element of $x$, 
and by $\Vert x\Vert$ the Euclidean norm of $x$.
Recall that \mbox{$\Vert x\Vert=\sqrt{x_1^2+x_2^2+\cdots+x_n^2}$}. The symbols $I_n$ and $0_{n\times{m}}$ denote the identity and zero matrices in $\R^{n\times{n}}$ and $\R^{n\times{m}}$, respectively. A continuous function \mbox{$\gamma:\mathbb{R}_{0}^{+}\rightarrow\mathbb{R}_{0}^{+}$}, is said to belong to class $\mathcal{K}$ 
if it is strictly increasing and \mbox{$\gamma(0)=0$}; $\gamma$ is said to belong to class $\mathcal{K}_{\infty}$ if \mbox{$\gamma\in\mathcal{K}$} 
and \mbox{$\gamma(r)\rightarrow\infty$} as $r\rightarrow\infty$. 
A continuous function \mbox{$\beta:\mathbb{R}_{0}^{+}\times\mathbb{R}_{0}^{+}\rightarrow\mathbb{R}_{0}^{+}$} is said to belong to class $\mathcal{KL}$ if, 
for each fixed $s$, the map $\beta(r,s)$ belongs to class $\mathcal{K}_{\infty}$ with respect to $r$
and, for each fixed nonzero $r$, the map $\beta(r,s)$ is decreasing with respect to $s$ and $\beta(r,s)\rightarrow0$ as \mbox{$s\rightarrow\infty$}. 

In this paper, we focus on \textit{linear} control systems given by the differential equation:
\begin{align}
\label{control_system}
\left\{\begin{array}{l}\dot\xi=A\xi+B\upsilon+\overline{B}\omega,\\
\eta=C\xi+\nu,\end{array}\right.
\end{align}
where, for any $t\in\R$, $\xi(t)\in\R^n$, $\upsilon(t)\in\R^m$, $\omega(t)\in\R^q$, $\eta(t)\in\R^p$, and $A$, $B$, $\overline{B}$, and $C$ are matrices of appropriate dimensions. 
The curve $\xi:\R\rightarrow\R^n$ is a \textit{trajectory} of (\ref{control_system}) if there exist curves $\upsilon:\R\rightarrow\R^m$ and $\omega:\R\rightarrow\R^q$ such that the time derivative of $\xi$ satisfies (\ref{control_system}). In the rest of the paper, we assume that all curves $\upsilon$ and $\omega$ have some regularity assumptions, guaranteeing existence and uniqueness of the solutions of (\ref{control_system}). Note that $\upsilon$, $\omega$, $\eta$, and $\nu$ denote control input, disturbance, output of the system and measurement noise, respectively. We assume that $\omega(t)$ and $\nu(t)$, for any $t\in\R$, are zero-mean Gaussian noise processes (uncorrelated from each other). For all curves $\omega$, we also write $\xi_{x\upsilon}(t)$ to denote the points reached at time $t$ under the input $\upsilon$ from initial condition $x=\xi_{x\upsilon}(0)$. 

To describe the mismatch between the controller specifications and its software implementations 
such as digital sampling and finite precision arithmetic, which is the focus of this paper, 
we consider the discrete-time version of (\ref{control_system}), as follows:
\begin{equation}
\label{control_system1}
\left\{\begin{array}{l}x[r+1]=A_\tau x[r]+B_\tau u[r]+\overline{B}_\tau d[r]+e_s,\\
\hspace{5.8mm}y[r]=Cx[r]+v[r],\end{array}\right.
\end{equation}
where the matrices $A_\tau$, $B_\tau$, and $\overline{B}_\tau$ are given by:
\begin{align}\nonumber
A_\tau=\textsf{e}^{A\tau},~~~~~B_\tau=\int_{r\tau}^{(r+1)\tau}\textsf{e}^{A(\tau-t)}Bdt,~~~~~\overline{B}_\tau=\int_{r\tau}^{(r+1)\tau}\textsf{e}^{A(\tau-t)}\overline{B}dt,
\end{align}
and $\tau$ is the sampling time. The function $\textsf{e}^{At}$, for any $t\in\R$, denotes the matrix function defined by the convergent series:
\begin{equation}\nonumber
\textsf{e}^{At}=I_{n}+At+\frac{1}{2!}A^2t^2+\frac{1}{3!}A^3t^3+\cdots,
\end{equation}
where \textsf{e} is Euler's constant. The signals $x$, $u$, $d$, $y$, and $v$ describe the exact value of the signals $\xi$, $\upsilon$, $\omega$, $\eta$, and $\nu$, respectively, at the sampling instants $0,\tau,2\tau,3\tau,\ldots$. Mathematically, we have:
\begin{align}\nonumber
x[r]=\xi(r\tau),~u[r]=\upsilon(r\tau),~d[r]=\omega(r\tau),~y[r]=\eta(r\tau),~v[r]=\nu(r\tau),~~\forall{r}\in\N_0.
\end{align}
The term $e_s$ in (\ref{control_system1}) is the sampling error. It can be shown that by sampling sufficiently fast, the error $e_s$ can be made arbitrarily small \cite{chen}. 
Since typical embedded controller implementations use sampling time in the range of milliseconds to microseconds, we will 
make the assumption that quantization errors dominate the sampling errors, and 
assume that $e_s=0$. 
 
We assume that only output of the system $y$ is measurable and not the full state $x$. 
Hence, a (proportional) {\em feedback} $K:\R^n\rightarrow\R^m$ defines the input $u[r]=-K\widehat{x}[r]$ based on an estimation $\widehat{x}$ of the state $x$. 
As explained in \cite{joao}, the estimation $\widehat{x}$ can be constructed using the observer dynamic:
\begin{equation}\label{observer}
\left\{\begin{array}{l}\widehat{x}[r+1]=A_\tau\widehat{x}[r]+B_\tau{u}[r]+L\left(y[r]-\widehat{y}[r]\right),\\
\hspace{5.6mm}\widehat{y}[r]=C\widehat{x}[r],\end{array}\right.
\end{equation}
where $\widehat{y}$ should be viewed as an estimate of $y$ and the linear map $L:\R^p\rightarrow\R^n$ is called an observer gain. 
By applying the feedback $u[r]=-K\widehat{x}[r]$ and combining the dynamics of control system in (\ref{control_system1}) and 
observer in (\ref{observer}), one obtains:
\begin{align}\label{overall0}
\left\{\begin{array}{l}x[r+1]=A_\tau x[r]-B_\tau K\widehat{x}[r]+\overline{B}_\tau d[r],\\\widehat{x}[r+1]=(A_\tau-B_\tau K-LC)\widehat{x}[r]+LCx[r]+Lv[r].\end{array}\right.
\end{align}

As shown in \cite{anta}, using a fixed-point implementation of the feedback gain as well as the observer dynamic, 
one gets the following overall dynamic:
\begin{align}\label{overal}
\left\{\begin{array}{l}x[r+1]=A_\tau x[r]-B_\tau K\widehat{x}[r]+\overline{B}_\tau d[r]+B_\tau e_{q2},\\\widehat{x}[r+1]=(A_\tau-B_\tau K-LC)\widehat{x}[r]+LCx[r]+Lv[r]+e_{q1},\end{array}\right.
\end{align}
where $e_{q1}$ and $e_{q2}$ are quantization errors in observer dynamic and feedback gain codes, respectively. 
Now, one can rewrite the control system in (\ref{overal}) as follows:
\begin{equation}\label{overal1}
w[r+1]=Gw[r]+H_1e_1[r]+H_2e_2[r],
\end{equation}
with: 
\begin{equation}\nonumber
w=\left[\begin{array}{l}x\\\widehat{x}\end{array}\right],~~e_1=\left[\begin{array}{l}d\\v\end{array}\right],~~e_2=\left[\begin{array}{l}e_{q1}\\e_{q2}\end{array}\right],
\end{equation}
and:
\begin{align}\nonumber
G=\left[\begin{array}{cc}A_\tau&-B_\tau{K}\\LC&A_\tau-B_\tau{K}-LC\end{array}\right],~H_1=\left[\begin{array}{cc}\overline{B}_\tau&0_{n\times{p}}\\0_{n\times{q}}&L\end{array}\right],~H_2=\left[\begin{array}{cc}0_{n\times{n}}&B_\tau\\I_{n}&0_{n\times{m}}\end{array}\right].
\end{align}
Since states of the control system (\ref{control_system}) are bounded physical quantities, such as 
temperature, pressure, position, velocity, acceleration and so on, their estimations and the output of the control system are bounded quantities as well. Hence, in the rest of the paper and without loss of generality, we assume that $y\in{Y}$, and $\widehat{x}\in\widehat{X}$, where $Y\subset\R^p$, and $\widehat{X}\subset\R^n$ are compact.

\subsection{Stability of perturbed systems}

Here, we recall the notion of uniform global asymptotic stability with respect to a set, 
presented in \cite{lin}. 

\begin{definition}[\cite{lin}]
\label{UGAS}
A control system of the form (\ref{control_system}) is uniformly globally asymptotically stable (UGAS) 
with respect to a set $\mathcal{A}$ if there exists a $\mathcal{KL}$ 
function $\beta$ such that for any $t\in\R_0^+$, any $x\in\R^n$, any control input $\upsilon:\R_0^+\rightarrow\mathsf{D}_1\subseteq\R^m$, and for all possible disturbances $\omega:\R_0^+\rightarrow\mathsf{D}_2\subseteq\R^q$, where $\mathsf{D}_1$, and $\mathsf{D}_2$ are compact sets, the following condition is satisfied:
\begin{equation}\label{GAS}
\Vert\xi_{x\upsilon}(t)\Vert_\mathcal{A}\leq\beta(\Vert{x}\Vert_\mathcal{A},t),
\end{equation} 
where the {\em point-to-set distance} $\Vert{x}\Vert_{\mathcal{A}}$ is defined by \mbox{$\Vert{x}\Vert_\mathcal{A}=\inf_{y\in\mathcal{A}}\Vert{x}-y\Vert$}. 
\end{definition}

When the set $\mathcal{A}$ is a singleton $\set{x_0}$, we speak of an asymptotically stable equilibrium
point $x_0$ rather than a UGAS set. The notion of UGAS for discrete-time control systems is obtained from Definition \ref{UGAS} by replacing $t\in\R_0^+$ with $r\in\N_0$.

We recall the following result describing how stability properties are affected by additive disturbances.
\begin{proposition}[\cite{anta}]
\label{prop:anta}
Consider the discrete-time linear system:
\begin{equation}\nonumber
x[r+1]=Ax[r],
\end{equation}
and assume that the origin is an asymptotically stable equilibrium point. Then, for any signal $d:\N_0\rightarrow\R^m$ satisfying $\left\Vert{d}[r]\right\Vert\leq b(d)$ for any $r\in\N_0$ and some constant $b(d)\in\R_0^+$, the system:
\begin{equation}\label{control_system4}
x[r+1]=Ax[r]+Bd[r],
\end{equation}
is UGAS with respect to the set:
\begin{equation}\nonumber
\mathcal{A}=\left\{x\in\R^n\,\,|\,\,\Vert{x}\Vert\leq\gamma b(d)\right\},
\end{equation} 
where $\gamma$ is given by:
\begin{equation}
\gamma=\max_{\theta\in[0,~2\pi[}\left\Vert\left(\textsf{e}^{i\theta}I_{n}-A\right)^{-1}B\right\Vert,
\end{equation}
with $i=\sqrt{-1}$. Moreover, the output $y=Cx$ is guaranteed to converge to the set:
\begin{equation}
\mathcal{A}_y=\left\{y\in\R^p\,\,|\,\,\Vert{y}\Vert\leq\gamma_y b(d)\right\},
\end{equation}  
with:
\begin{equation}\nonumber
\gamma_y=\max_{\theta\in[0,~2\pi[}\left\Vert C\left(\textsf{e}^{i\theta}I_{n}-A\right)^{-1}B\right\Vert.
\end{equation}
\end{proposition}

In control theory, $\gamma_y$ is known as the {\em $\mathcal{L}_2$ gain} of the control system in (\ref{control_system4}) with the output $y=Cx$.
The following proposition follows from Proposition~\ref{prop:anta} and describes the stability properties of linear control systems in (\ref{overal1}) with 
respect to disturbance, measurement noise, and implementation errors in the feedback gain and observer dynamic. 

\begin{proposition}
\label{prac_stability1}
Consider the discrete-time linear system in (\ref{overal1}). Then for any input $e_1$ and $e_2$ satisfying $\Vert{e_1}[r]\Vert \leq b(e_1)$ and $\Vert{e_2}[r]\Vert \leq b(e_2)$ for any $r\in\N_0$ and some
constants $b(e_1),b(e_2)\in \mathbb{R}^+_0$, the system is UGAS with respect to the set:
\begin{equation}\nonumber
\mathcal{A}=\left\{x\in\R^n\,\,|\,\,\Vert{x}\Vert\leq\gamma_1b(e_1)+\gamma_2b(e_2)\right\},
\end{equation} 
where $\gamma_1$ and $\gamma_2$ are given by: 
\begin{align}\nonumber
\gamma_j=\max_{\theta\in[0,~2\pi[}\left\Vert\left(\textsf{e}^{i\theta}I_{2n}-G\right)^{-1}H_j\right\Vert,~~\text{for}~j=1,2,
\end{align}
with $i=\sqrt{-1}$. Moreover, the output $\R^p\ni y=\left[C~~0_{p\times{n}}\right]w$ is guaranteed to converge to the set:
\begin{equation}\label{prac_stable_reg}
\mathcal{A}_y=\left\{y\in\mathbb{R}^p\mid\Vert y\Vert\leq\gamma_{1y}b(e_1)+\gamma_{2y}b(e_2)\right\},
\end{equation}
where $\gamma_{1y}$ and $\gamma_{2y}$ are given by: 
\begin{align}\label{l2_gain}
\gamma_{jy}=\max_{\theta\in[0,~2\pi[}\left\Vert\left[C~~0_{p\times{n}}\right]\left(\textsf{e}^{i\theta}I_{2n}-G\right)^{-1}H_j\right\Vert,~~\text{for}~j=1,2.
\end{align}
\end{proposition}

The error vector $e_1$ includes disturbance and measurement noise, depending for example on the environment and the quality of the sensors collecting measurements. Hence, the controller designer does not have any control on the value of $b(e_1)$. However, one can reduce the amount of $\gamma_{1y}$ by appropriately choosing gains $K$ and $L$. On the other hand, one can reduce the amount of not only $\gamma_{2y}$ but also $b(e_2)$ by appropriately choosing gains $K$ and $L$. We use Proposition~\ref{prac_stability1} in the following way.
Given a feedback gain $K$ and an observer gain $L$, we compute $\mathcal{L}_2$ gains $\gamma_{1y}$ and $\gamma_{2y}$ and an upper bound $b(e_2)$ on the implementation error $e_2$. Then the output of the controlled system (with
implementation error) must converge to set $\mathcal{A}_y$ in (\ref{prac_stable_reg}).
We show later that appropriate choices of gains $K$ and $L$ can shrink the size of the 
set $\mathcal{A}_y$ and hence, provide a tighter bound on the set 
to which the output of the system converges.

\subsection{LQR-LQG performance}
In addition to asymptotic stability, controller designers also consider the {\em performance} of the controller,
that is, of the controllers ensuring asymptotic stability of the origin, one desires the controller that minimizes
a given cost function.
A common approach for optimal output feedback controller are the {\em linear quadratic regulator} (LQR) and {\em linear quadratic Gaussian} (LQG). 
The LQR cost function to be minimized is given by:
\begin{equation}
\label{LQR_cost}
J_{LQR}=\sum_{r=0}^{+\infty}\left\{x[r]^TQx[r]+u[r]^TRu[r]\right\},
\end{equation}
for some chosen weight matrices $Q$ and $R$ that are positive definite and of appropriate dimensions. 

The LQG cost function to be minimized is given by:
\begin{equation}\label{LQG}
J_{LQG}=\lim_{r\rightarrow+\infty}\text{E}\left[\Vert e[r]\Vert^2\right],
\end{equation}
where $\text{E}$ stands for expected value and $e$ is the estimation error for the control system in (\ref{overall0}) whose dynamic is given by:
\begin{equation}\label{stochastic_dyn}
e[r+1]=x[r+1]-\widehat{x}[r+1]=(A_\tau-LC)e[r]+\overline{B}_\tau d[r]-Lv[r].
\end{equation}
As mentioned before, $d$ and $v$ are assumed to be zero-mean Gaussian noise process (uncorrelated from each other) with covariance matrices:
\begin{equation}\label{spectrum}
\text{E}\left(d[r]d[r]^T\right)=\widehat{Q},~\text{E}\left(v[r]v[r]^T\right)=\widehat{R},~\forall r\in\N_0,
\end{equation}
where $\widehat{Q}$ and $\widehat{R}$ are some positive semi-definite matrices of appropriate dimensions.

A standard control-theoretic construction rewrites the cost function (\ref{LQR_cost}) as $J_{LQR}=x[0]^T S(K)x[0]$, where $u=-Kx$, and $S(K)\in\mathbb{R}^{n\times n}$ is a positive definite matrix that is the unique solution for $S$ to the Lyapunov equation:
\begin{equation}\label{lyapunov}
\left(A_\tau-B_\tau K\right)^TS\left(A_\tau-B_\tau K\right)-S+Q+K^TRK=0,
\end{equation}
where $K$ is a controller making $A_\tau-B_\tau K$ Hurwitz.\footnote{
We call the matrix $A_\tau-B_\tau K$ Hurwitz if its eigenvalues are inside the unit circle, centered at the origin.} 
See \cite{joao} for detailed information. Additionally, we have
\begin{equation}
\lambda_{\min}(S(K))\Vert{x[0]}\Vert^2\leq{J_{LQR}}\leq\lambda_{\max}(S(K))\Vert{x[0]}\Vert^2,
\end{equation}
where $\lambda_{\min}(S(K))\in\R^+$ and $\lambda_{\max}(S(K))\in\R^+$ are minimum and maximum eigenvalues of $S(K)$, respectively. 
Therefore, $J_{LQR}$ can be minimized for all possible choice of initial conditions by just minimizing the maximum eigenvalue of $S(K)$. Note that since $S$ is a positive definite matrix, its maximum eigenvalue is equal to its induced 2 norm\footnote{We recall that induced 2 norm of a matrix $A\in\R^{n\times{m}}$ is given by: $\Vert{A}\Vert=\sqrt{\lambda_{\max}\left(A^TA\right)}$.} $\Vert{S}\Vert$.  

Similarly, the cost function (\ref{LQG}) can be rewritten as $J_{LQG}=\Vert{P(L)}\Vert$, where $P(L)\in\mathbb{R}^{n\times n}$ is a positive definite matrix that is the unique solution for $P$ to the Lyapunov equation:
\begin{equation}\label{lyapunov1}
\left(A_\tau-LC\right)P\left(A_\tau-LC\right)^T-P+\overline{B}_\tau\widehat{Q}\overline{B}_\tau^T+L\widehat{R}L^T=0,
\end{equation}
where $L$ is an observer gain making $A_\tau-LC$ Hurwitz. See \cite{joao} for more detailed information. Therefore, $J_{LQG}$ can be minimized by just minimizing $\Vert P(L)\Vert$.

Note that the optimal feedback $u=-Kx$ minimizing the LQR cost in (\ref{LQR_cost}) is computed using the deterministic dynamic: $$x[r+1]=A_\tau x[r]+B_\tau u[r].$$ On the other hand, the optimal gain $L$ minimizing the LQG cost in (\ref{LQG}) is computed using the stochastic dynamic in (\ref{stochastic_dyn}). Thanks to the separation principle for linear control systems \cite{joao}, one concludes that the overall closed loop system in (\ref{overall0}) is $UGAS$ even though the gains $K$ and $L$ are designed separately.

\subsection{The effect of errors}

\myparagraph{Example}
We now present a simple motivating example showing how different choice of
controllers result in different steady state errors due to their fixed-point implementations, 
yet providing approximately the same LQR-LQG performance. 
Consider a simple physical model of a bicycle, borrowed from \cite{astrom}. 
The dynamics of the system is given by:
\begin{align}
\label{bicycle}
\left\{\begin{array}{l}\left[{\begin{array}{c}
\dot\xi_1\\
\dot\xi_2\\
 \end{array}}\right]=\left[ {\begin{array}{cc}
0&\frac{g}{h} \\
1&0\\
 \end{array}}\right]\left[{\begin{array}{c}
\xi_1\\
\xi_2\\
 \end{array}}\right]+\left[ {\begin{array}{c}
1\\
0\\
 \end{array}}\right]\left(\upsilon+\omega\right),\\ \hspace{7.8mm}\eta=\left[\frac{av_0}{bh}~~\frac{v_0^2}{bh}\right]\left[{\begin{array}{c}
\xi_1\\
\xi_2\\
 \end{array}}\right]+\nu,\end{array}\right.
\end{align}
where $\xi_1$ is the steering angular velocity, $\xi_2$ is the steering angle, $\eta$ is the role angle, $\upsilon$ is the torque applied to the handle bars, 
$g=9.8m/s^2$ is the acceleration due to gravity, $h=1.5m$ is the height of the center of mass, 
$v_0=2m/s$ is the velocity of the bicycle at the rear wheel, 
$a=0.5m$ is the distance of the center of mass from a vertical line through the contact point of the rear wheel and $b=1m$ is the wheel base. 

The control objective is to design a feedback gain $K\in\R^{1\times2}$ and an observer gain $L\in\R^{2\times{1}}$ such that the feedback control law $u=-K\widehat{x}$, 
where \mbox{$\widehat{x}=[\widehat{x}_1,~\widehat{x}_2]^T$} is the state of the observer in (\ref{observer}), 
makes the closed loop system UGAS. 
By choosing the matrices $Q=I_2$ and $R=1$ inside the LQR cost function and $\widehat{Q}=1$ and $\widehat{R}=1$ in (\ref{spectrum}), 
the feedback and observer gains minimizing the LQR and LQG costs are given by $K_1=[5.1538,~12.9724]$, and $L_1=[0.0317,~0.0118]^T$, respectively. 
Consider a second pair of feedback and observer gains given by $K_2=[3.0253,~12.6089]$ and $L_2=[0.0132,~0.1021]^T$. 
For the initial condition $x=(0.2,~0.2)^T$, the value of the LQR cost function is $264.1908$ for feedback gain $K_1$ and 
$284.1578$ for $K_2$. 
Moreover, the value of the LQG cost function is $0.0229$ for observer gain $L_1$ and $0.0246$ for $L_2$. 
That is, the gains $K_2$ and $L_2$ give cost functions about 7\% greater than the optimal gains $K_1$ and $L_1$.

\begin{figure}
  \centering\includegraphics[width=13cm]{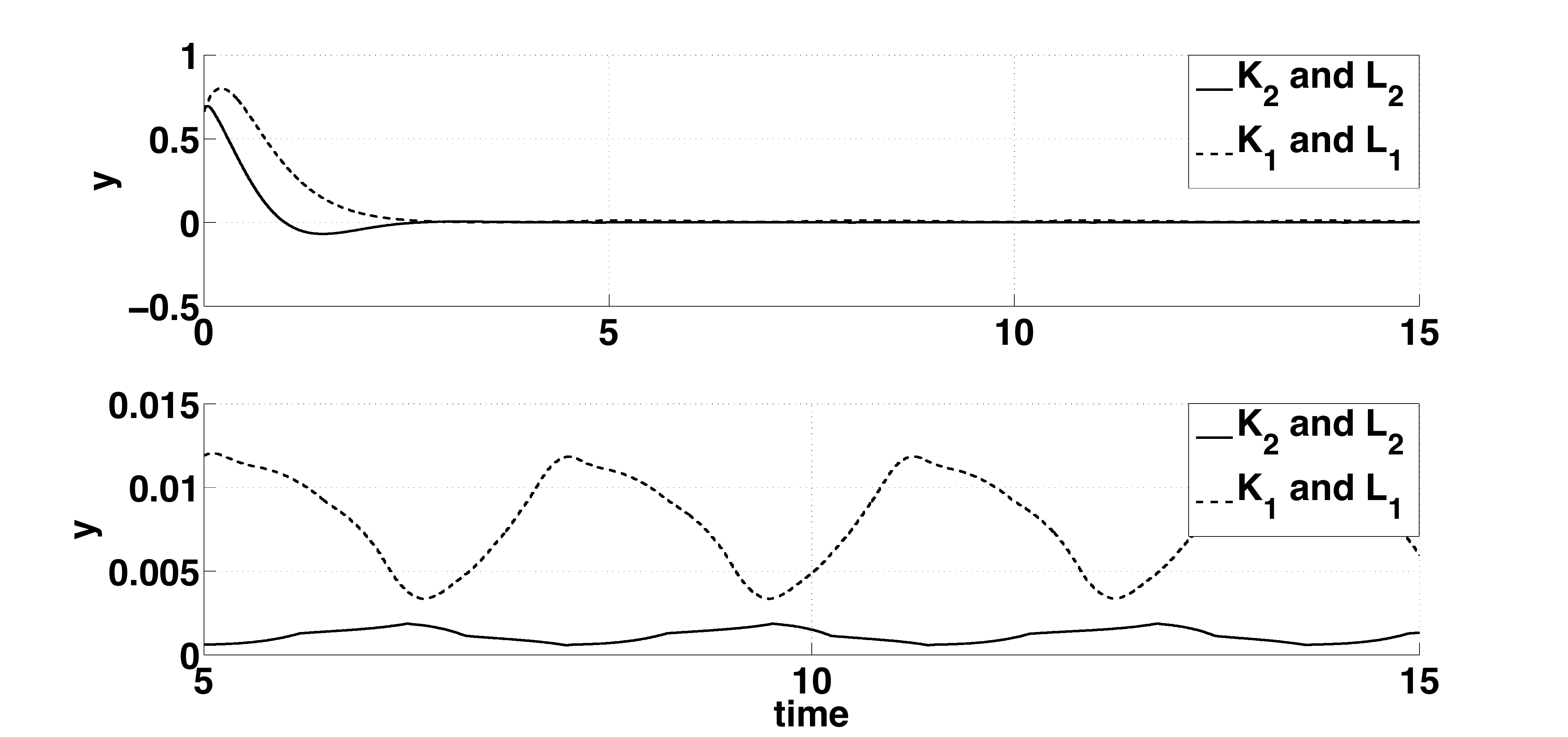}
  \caption{Evolution of the output $y$ with initial state $(0.2,~0.2)^T$ for the pair of gains $K_1$, $L_1$ and $K_2$, $L_2$ using 16-bit implementation.}
\label{fig1}
\end{figure}

We now show how different choice of feedback and observer gains result in different fixed-point implementation errors.
For now, let us assume that $\omega(t)=0$ and $\nu(t)=0$, for any $t\in\R_0^+$. 
In Figure \ref{fig1}, we show the output of the closed-loop system starting from the 
initial condition $x = (0.2,~0.2)^T$, 
when the feedback gain and observer dynamic are implemented using 16-bit fixed-point representation. 
As can be observed from Figure~\ref{fig1}, 
the output of the controlled system 
does not converge to the equilibrium point at the origin because 
of the fixed-point implementation error in the controllers. Furthermore, the practical stability region using gains $K_2$ and $L_2$ is much smaller than the one using gains $K_1$ and $L_1$.

Using bounds on implementation errors for the two controllers (described in Section~\ref{error_comp})
and Proposition~\ref{prac_stability1}, we can prove that the output of the system with feedback and observer gains $K_1$ and $L_1$ 
(resp.\ $K_2$ and $L_2$) converges to a ball centered at the origin with radius $0.5486$ (resp.\ $0.0513$), 
whenever the output of the system and the state of the observer take values in the interval $[-1,~1]$ and the feedback gain and observer dynamic are implemented 
using 16-bit fixed-point implementation. 
As can be seen, given a 16-bit implementation, feedback and observer gains $K_2$ and $L_2$ may be preferred to gains $K_1$ and $L_1$ 
because they have guaranteed bounds on practical stability region that is 10 times smaller than gains $K_1$ and $L_1$ and provide approximately similar performance. 
If one considers the effect of disturbance and measurement noise, it can be proved that the output of the system with feedback and observer gains 
$K_1$ and $L_1$ (resp.\ $K_2$ and $L_2$) converges to a ball centered at the origin with radius $5.0489b(e_1)+0.5486$ (resp. $2.5341b(e_1)+0.0513$),
where $b(e_1)$ is an upper bound on the size of the vector $e_1$ introduced in (\ref{overal1}).

\myparagraph{Optimization objectives}
The above example suggests that the control design should optimize for the following objectives: 
the LQR and the LQG costs for performance, error caused by disturbance and measurement noise, 
and the implementation error given by a fixed-precision encoding. 
Accordingly, we define a cost function that is weighted sum of the four factors:
\begin{equation}\label{cost1}
\mathcal{J}(K,L) = w_1\frac{\Vert S(K)\Vert}{\Vert S^*\Vert}+w_2\frac{\Vert P(L)\Vert}{\Vert P^*\Vert}+w_3\frac{\gamma_{1y}}{\gamma^*_{1y}}+w_4\frac{\gamma_{2y}b(e_2)}{\gamma^*_{2y}b^*(e_2)},
\end{equation}
where $w_1,\ldots,w_4$ are weighting factors, $S^*$ and $P^*$ are matrices, computed from Lyapunov equations in (\ref{lyapunov}) and (\ref{lyapunov1}) using standard LQR and LQG gains ($K_{LQR}$ and $L_{LQG}$), 
$\gamma_{1y}$ and $\gamma_{2y}$ (resp. $\gamma^*_{1y}$ and $\gamma^*_{2y}$) are the $\mathcal{L}_2$ gains in (\ref{l2_gain}) using feedback and observer gains $K$ and $L$ 
(resp. $K_{LQR}$ and $L_{LQG}$) and $b(e_2)$ (resp. $b^*(e_2)$) is the bound on the implementation error of given feedback and observer gains 
$K$ and $L$ (resp. $K_{LQR}$ and $L_{LQG}$). 
Minimizing the terms $\gamma_{1y}$ and $\gamma_{2y}b(e_2)$ inside \eqref{cost1} results in a tighter bound 
on the set $\mathcal{A}_y$ in Proposition~\ref{prac_stability1}. 
Since the four factors in (\ref{cost1}) have different scales, we normalized them by their values using the standard gains 
$K_{LQR}$ and $L_{LQG}$. 
The designer can choose $w_1,\ldots,w_4$ based on the priorities on LQR and LQG performances and steady state error.
Our objective is to find feedback and observer gains that minimize the cost function $\mathcal{J}$.

We focus on implementation errors arising out of fixed-precision arithmetic. The bound $b(e_2)$ is computed using the strategy, explained in Section \ref{error_comp}. 
Note that the cost function $\mathcal{J}$ is not necessarily convex with respect to the feedback and observer gains $K$ and $L$. 
Therefore, the proposed design strategy cannot be formulated as a convex optimization problem. We use a heuristic stochastic optimization approach
to find feedback and observer gains $K$ and $L$ minimizing $\mathcal{J}$. 

In our exposition, we consider the plant model to be precise, and only consider quantization effects as the source of error.
Our methodology can consider both additive and multiplicative uncertainties in the plant model as well \cite{green}. 
We can take those uncertainties into account by adding appropriate extra terms to the cost function in (\ref{cost1}) 
using the results provided in \cite{majid1,majid2}. 
We omit the details for simplicity.

\section{Computing Quantization Error }\label{error_comp}

In this section we show how to compute a bound on the fixed-point implementation error for given feedback and observer gains $K$ and $L$. 
We assume that the outputs of the controlled system and the state of the observer are restricted to compact subsets $Y\subset\R^p$ and $\widehat{X}\subset\R^n$, respectively. 

\subsection{Best fixed-point implementation}
A {\em fixed-point representation} of a real number is a triple $\tuple{s,n,m}$ 
consisting of a {\em sign bit} $s\in\set{\mathtt{s},\mathtt{u}}$ (for {\em signed} and {\em unsigned}),
a {\em length} $n\in\nats$, and a {\em length of the fractional part} $m\in\nats$.
The length of the integer part is $n - m -1$. 
Intuitively, a real number is represented using $n$ bits, of which $m$ bits are used to store
the fractional part.
Clearly, the largest integer portion has to fit in $n-m-1$ bits.

A variable with a fixed-point type is represented as an integer. 
We associate an integer variable $\hat{x}$ with the fixed-point representation of a real variable $x$.
An integer variable $\hat{x}$ that represents a fixed-point variable with type  $\tuple{\mathtt{u},n,m}$
can be interpreted as the rational number $2^{-m} \hat{x}$.
We deal with a signed number by separately tracking the sign and the magnitude,
performing the operations on the magnitudes using unsigned arithmetic, and
finally putting the appropriate sign bits back.

An operation using real arithmetic may have different fixed-point implementation depending on
how many bits are allocated to hold the integer part and the fraction part of the variables. 
Allocating fewer number of bits than required to hold the integer part may lead to overflow. 
On the other hand, if more than the required number of bits are allocated to the integer part, 
the quantization error increases due to assigning few bits to the fractional part. 
When we compare fixed-point implementation of different controllers, we first synthesize the best 
possible implementation of a controller. 

Let us fix the number of bits to be $n$ for the implementation of a controller. 
With $n$ bits, let $b$ be the upper bound on the quantization error in a fixed-point implementation $I$ of a controller
for a given range for the inputs. 
The fixed-point implementation $I$ is the {\em best implementation} if  there 
does not exist another implementation $I'$ using $n$ bits, 
for which the upper bound on the quantization error is $b'$ and $b' < b$.

If the ranges of the variables in the real arithmetic computation can be computed exactly,  
it is possible to synthesize the best fixed-point implementation. In the best fixed-point implementation,
the number of bits allocated to the integer part is just enough to hold the integer part of any value in that range.
For example, if the range of a variable is \mbox{[-35.55,~48.72]}, the datatype for the variable in the best
16-bit fixed-point representation is $\tuple{1,16, 9}$.  

The range computation problem
of variable $y$ in an operation 
$y = f(x_1, \ldots , x_n)$
involves solving a maximization and a minimization
problem where $f$ is the objective function and the ranges on $x_1, \ldots, x_n$ form the set of constraints.
If the function $f$ is convex, the range of $y$ can be computed exactly, and also it is straight-forward to find 
the best fixed-point implementation for the operation.

\subsection{Error bound computation}

We apply mixed-integer linear programming based optimization technique to find out the error bound
between a computation in real arithmetic and its best fixed-point implementation.
Suppose we have an arithmetic operation $s: a = b \ op \ c$, where $op \in \set{+,-,*}$. If $op = *$, then
either $b$ or $c$ is a constant. If $op = +$ or $op = -$, then $b$ and $c$ can both be variables. 
We associate an integer variable $\hat{x}$ with the fixed-point representation of a real variable $x$.
Let the range of the values for $a$ and $b$ and $c$
are $[l_a, u_a]$, $[l_b, u_b]$, and $[l_c, u_c]$, respectively. Let the fixed-point representation of 
$a$, $b$ and $c$ are $\tuple{\mathtt{s}, n_a,m_a}$, $\tuple{\mathtt{s}, n_b,m_b}$,  and $\tuple{\mathtt{s}, n_c,m_c}$, 
respectively. Let $b(e_b)$ and $b(e_c)$ be bounds on the quantization errors of $b$ and $c$, respectively.
The optimization problem to find the bound on the error is given by:
\begin{equation}
\label{opt_prob}
\begin{tabular}{rl}
maximize & $\left\vert a - 2^{-m_a} \hat{a}\right\vert$\\
Subject to & $l_a \le a \le u_a$\\
& $l_b \le b \le u_b$\\
& $\left\vert b - 2^{-m_b}*\hat{b}\right\vert \le b(e_b)$\\
& $\left\vert c - 2^{-m_c}*\hat{c}\right\vert \le b(e_c)$\\
& $a = b \ op \ c$ \\
& $\Phi(\mathsf{fp}(s))$
\end{tabular}
\end{equation}
where $\mathsf{fp}(s)$ is the fixed-point representation of the statement $s$
and 
$\Phi(\mathsf{s})$ denotes a logical formula that relates the inputs and outputs of the
fixed-point representation $\mathsf{s}$. Technically, $\Phi$ is the {\em strongest postcondition}~\cite{Winskel} of $\mathsf{s}$ with respect to\ $\true$. We compute $\Phi$ 
using an arithmetic encoding of a fixed-point computation~\cite{anta}. 
Here we illustrate the computation of the strongest
postcondition $\Phi$ using an example.

\smallskip
\noindent
{\bf Example.} Suppose we have the following arithmetic operation 
$$s: \ y = -7.2479 \ * \ x \ .$$ 
Assume the compact set for $x$ is [-1, 1]. The fixed-point expression corresponding to $s$ in the best fixed-point implementation is 
$$\mathsf{fp}(s) :  -\hat{y} = (-115 * \hat{x}) \gg 6 \ .$$
The strongest postcondition $\Phi(\mathsf{fp}(s))$ of $\mathsf{fp}(s)$ is given by:\\
\begin{center}
\begin{tabular}{rl}
$\Phi(\mathsf{fp}(s)) :=$ & $tmp = -115 * \hat{x} \ \wedge$ \\
& $tmp \ge  0 \rightarrow tmp1 = tmp \ \wedge$ \\
& $tmp < 0 \rightarrow tmp1 = -tmp \ \wedge$ \\
& $tmp1 = 2^6 * divisor + remainder \ \wedge$ \\ 
& $remainder \ge 0 \wedge remainder < 2^6 \ \wedge$ \\ 
& $tmp \ge 0 \rightarrow \hat{y} = divisor \ \wedge$ \\ 
& $tmp < 0 \rightarrow \hat{y} = - divisor \ $,
\end{tabular}
\end{center}
where $tmp$, $tmp1$, $divisor$, and $remainder$ are integer variables.

Depending on the arithmetic operation, we need to solve at most 4 instances of mixed integer linear
programming problem to solve the optimization problem in (\ref{opt_prob}), and the maximum among all of them gives the bound on the error in the
fixed-point implementation.

We use the above technique to compute the bound on the error in one operation in the 
fixed-point implementation of a gain. The implementation of a gain
involves a series of arithmetic operations. We compute the error bound for the output of one arithmetic operation
at a time. Let $s: a = b \ op \ c$ is an arithmetic operation in the implementation of a gain. In the arithmetic operation,
$b$ and $c$ may either be a constant, a state variable or a temporary variable which captures the result of some
previous operation. If $b$ (or $c$) represents a constant, and the fixed-point representation contains $m$ bits for the 
fraction part, then the error in the fixed point representation is bounded by $\frac{1}{2^m}$. If $b$ (or $c$) represents
a state variable, then the fixed-point datatype can be determined from the given compact set for the state, and the fixed-point
datatype can be determined accordingly. Then the error in the fixed-point representation is bounded by $\frac{1}{2^m}$, where
$m$ is the number of bits to represent the fraction part in the fixed-point datatype of the variable. If  $b$ (or $c$) is a temporary variable used to hold 
the result of an earlier computation, then the range and error bound for the variable is already known.

\section{Optimal Controller Synthesis}
We now describe our controller synthesis algorithm that minimizes the cost function~\eqref{cost1}
combining LQR and LQG performance, disturbance, measurement noise and implementation errors.
Since the cost function is non-convex, we use a stochastic local search technique.

\subsection{Particle swarm optimization (PSO)}\label{PSO}
PSO is a stochastic local search approach.
It maintains a set of potential solutions (called ``particles'')
in a compact $d$ dimensional search space $D=\prod_{j=1}^d [y_{\min}^j,y_{\max}^j]\subset\R^d$, minimizing a given cost function.
The particles move in this space according to their velocity.
Each particle, indexed by $i\in\N$, has a position $y_i\in\R^d$, changing between $y_{\min}$ and $y_{\max}$, and a velocity vector $v_i\in\R^d$, changing between some vectors $v_{\min}$ and $v_{\max}$. 
The terms $v_{\min}$ and $v_{\max}$ are often set to the maximum dynamic range of the variables on each dimension \cite{majid}: $-v_{\min}^j=v_{\max}^j=\vert y_{\max}^j-y_{\min}^j\vert$. 
Every particle remembers its own best position (i.e., the lowest value of the cost function achieved so far by this particle) in a 
vector $P_i\in\R^d$.
The best position with respect to the cost function among all of the particles 
so far is stored in a vector $P_g\in\R^d$.

PSO updates the positions and velocities of all particles iteratively.
The new velocity and position for particle $i$ are determined as:
\begin{align}
\label{velocity}
v_{i}^{l+1}=&w^lv_{i}^l+c_1r_1\left(P_i^l-y_{i}^l\right)+c_2r_2\left(P_g^l-y_{i}^l\right),\\\label{position}
y_{i}^{l+1}=&y_{i}^l+v_{i}^{l+1},
\end{align}
where the superscript $l$ denotes the iteration number, the subscript
$i=1,\ldots,N$ denotes the index of the particle, and $N$ is the number of particles. 
The constant $w^l$ in (\ref{velocity}) is updated using the inertia weight approach \cite{ebbesen} as the following:
\begin{equation}
w^l=\max\left\{w_{\min},w_{\max}-\frac{w_{\max}-w_{\min}}{l_{\max}}(l-1)\right\},
\end{equation} 
where $w_{\max}$ and $w_{\min}$ are adjusted to $1$ and $\frac{c_1+c_2}{2}-1$ and $l_{\max}$ is the maximum number of iterations. 
The constants $c_1$ and $c_2$ in (\ref{velocity}) are the acceleration constants, influencing the convergence speed of particles toward its own and global best positions
and set to $0.5$ and $1$, respectively \cite{ebbesen}. 
The constants $r_1$ and $r_2$ in (\ref{velocity}) are uniformly distributed random numbers on the interval $[0,1]$.

\subsection{Overall algorithm} \label{algorithm}

The PSO algorithm is used to search for feedback and observer gains $K\in\R^{m\times{n}}$ and $L\in\R^{n\times{p}}$ 
for the control system (\ref{overal}), minimizing (\ref{cost1}). 
Note that a particle in PSO represents a feedback and an observer gain $K$ and $L$, respectively, moving in an $m\times{n}+n\times{p}$ dimensional search space. 
To discard those gains that make the controlled system unstable, we penalize unstable gains by including a penalty term $\widetilde{P}$ in the cost function
such that $\widetilde{P} = 0$ if $A_\tau - B_\tau K$ and $A_\tau-LC$ are Hurwitz and $\widetilde{P} = +\infty$ otherwise. 
The cost function for PSO is then $F(K,L) = \mathcal{J}(K,L) + \widetilde{P}(K,L)$.

The design steps can be summarized as the following:
\begin{itemize}
\item[(1)] Initialize positions of $N$ feedback gains $K_i$ and observer gains $L_i$ by $K_{LQR}$ and $L_{LQG}$, respectively, and uniformly randomly initialize their velocities , $i=1,\ldots,N$, where $N$ is the number of particles.
\item[(2)] Given any initial feedback gain $K_i$ and observer gain $L_i$, compute the cost function $F(K_i,L_i)$.
To compute $\widetilde{P}$, check if $A_\tau - B_\tau K$ and $A_\tau-LC$ are Hurwitz.
There are some steps to compute $\mathcal{J}$.
First compute $S(K_i)$ and $P(L_i)$ by solving the Lyapunov equations~\eqref{lyapunov} and (\ref{lyapunov1}), respectively, and find their induced 2 norm.
Second, compute the $\mathcal{L}_2$ gains $\gamma_{1y}$ and $\gamma_{2y}$. Third, compute $b(e_2)$ by solving the optimization problems from Section~\ref{error_comp}.
\item[(3)] Compare $F(K_i,L_i)$ to its own best position $P_i$ so far and the global best position $P_g$ so far.
If $F(K_i,L_i)$ is less than the previous personal best (resp.\ the global best), update the best position (resp.\ the global best) to $K_i$ and $L_i$.
\item[(4)] Modify the velocity and position of each pair $K_i$ and $L_i$ according to (\ref{velocity}) and (\ref{position}).
\item[(5)] If the number of iterations, denoted by $l$, reaches the maximum, denoted by $l_{\max}$, or the value of $F$ does not change for the global best position $P_g$ for 50 consecutive iterations up to error $10^{-6}$ then go to Step (6), otherwise go to Step (2);
\item[(6)] The latest $P_g$ is the optimal controller.
\end{itemize}

\section{Extension: PID Controllers}\label{PID}

PID controllers are a common class of controllers in many industries, such as automotive, power systems, servomotors, and so on. 
We now extend the analysis of Section \ref{preliminaries} to PID controllers. 
A PID controller generalizes a proportional feedback controller, and includes three terms: proportional, integrator, and differentiator. 
For an input $\upsilon$, the output $\eta$ of the PID controller is computed as follows:
\begin{equation}\label{PID_continuous}
\eta(t)=K_P\upsilon(t)+K_I\int_0^t\upsilon(s)ds+K_D\frac{d\upsilon(t)}{dt},~~\forall t\in\R_0^+,
\end{equation}   
where $K_P$, $K_I$, and $K_D$ are called proportional, integrator, and differentiator gains, respectively. 
To describe the mismatch between the PID specifications and its software implementation, 
we consider the discrete-time version of (\ref{PID_continuous}). 
A common way of discretizing an integrator is based on the trapezoidal approximation. 
An integrator term:
\begin{equation}\nonumber
\eta(t)=\int_0^t\upsilon(s)ds,~~\forall t\in\R_0^+,
\end{equation}  
can be discretized as follows:
\begin{equation}\label{trapezoidal}
y[r+1]=y[r]+\frac{\tau}{2}\left(u[r+1]+u[r]\right),~~\forall r\in\N_0,
\end{equation}  
where $\tau$ is the sampling time, $y[r]=\eta(r\tau)+e_1$ and $u[r]=\upsilon(r\tau)$, for any $r\in\N_0$. A common way of discretizing a differentiator, is based on the backward Euler method. A differentiator term:
\begin{equation}\nonumber
\eta(t)=\frac{d\upsilon(t)}{dt},~~\forall t\in\R_0^+,
\end{equation}  
can be discretized as follows:
\begin{equation}\label{BE}
y[r+1]=\frac{u[r+1]-u[r]}{\tau},~~\forall r\in\N_0,
\end{equation}  
where $y[r]=\eta(r\tau)+e_2$ and $u[r]=\upsilon(r\tau)$, for any $r\in\N_0$. By using fast sampling time assumption, we can ignore the errors $e_1$ and $e_2$ in the discretized versions of integrator and differentiator in comparison with quantization errors. To follow the same analysis as in Section \ref{preliminaries}, we need a state space realization of PID controller. By resorting to results in control classic \cite{kailath} and using the discretization rules in (\ref{trapezoidal}) and (\ref{BE}), the state space realization of discretized PID controller with input $\widehat{u}[r]$ and output $\widehat{y}[r]$ is obtained as follows:
\begin{eqnarray}\label{state_PID}
\left\{\begin{array}{l}\widehat{x}[r+1]=\widehat{A}\widehat{x}[r]+\widehat{B}\widehat{u}[r],\\
\hspace{6mm}\widehat{y}[r]=\widehat{C}\widehat{x}[r]+\widehat{D}\widehat{u}[r],\end{array}\right.
\end{eqnarray}
where
\begin{eqnarray}\nonumber
\widehat{A}=\left[ {\begin{array}{cc}
0&1 \\
0&1\\
 \end{array}}\right],~\widehat{B}=\left[ {\begin{array}{c}
0\\
1\\
 \end{array}}\right],~\widehat{C}=\left[\frac{K_D}{\tau}~~~K_I\tau-\frac{K_D}{\tau}\right],~\widehat{D}=\left(K_P+\frac{K_I\tau}{2}+\frac{K_D}{\tau}\right).
\end{eqnarray}
Without loss of generality, consider a single-input ($m=1$) single-output ($p=1$) discrete-time linear control system of the form:
\begin{eqnarray}\nonumber
\left\{\begin{array}{l}x[r+1]=Ax[r]+Bu[r],\\
\hspace{6mm}y[r]=Cx[r].\end{array}\right.
\end{eqnarray}
Since the input of the PID controller is equal to the negative of the output of the plant ($\widehat{u}=-y$) because of negative feedback and the output of the PID controller is equal to the input of the plant ($u=\widehat{y}$), one obtains:
\begin{eqnarray}
\left\{\begin{array}{l}x[r+1]=\left(A-B\widehat{D}C\right)x[r]+B\widehat{C}\widehat{x}[r],\\
\widehat{x}[r+1]=-\widehat{B}Cx[r]+\widehat{A}\widehat{x}[r].\end{array}\right.
\end{eqnarray}
Similar to what explained in Section \ref{preliminaries}, by fixed-point implementation of the PID controller, one gets the following overall dynamic:
\begin{eqnarray}\label{overall5}
\left\{\begin{array}{l}x[r+1]=\left(A-B\widehat{D}C\right)x[r]+B\widehat{C}\widehat{x}[r]+Be_{q2},\\
\widehat{x}[r+1]=-\widehat{B}Cx[r]+\widehat{A}\widehat{x}[r]+e_{q1},\end{array}\right.
\end{eqnarray}
where $e_{q1}$ and $e_{q2}$ are quantization errors in computing the PID controller. Now, we can use the same strategy, as explained in Subsection \ref{algorithm}, to design parameters $K_P$, $K_I$, and $K_D$ of PID controllers minimizing a performance-based cost function as well as the effect of quantization error. For example, one can consider:
\begin{equation}\label{cost2}
\mathcal{J}(K_P,K_I,K_D) = \frac{w_1}{\text{PM}}+\frac{w_2}{\text{GM}}+w_3\gamma(b(e_{q1})+b(e_{q2})),
\end{equation}
where PM and GM are phase and gain margins, $w_1,w_2,w_3$ are weighting factors, $\gamma$ is the $\mathcal{L}_2$ gain of the linear control system (\ref{overall5}) and $b(e_{q1})$ and $b(e_{q2})$ are the bounds on the implementation errors $e_{q1}$ and $e_{q2}$. Note that control over PM and GM guarantees robust stability of the closed-loop systems \cite{joao}. The phase and gain margins measure the system's tolerance to the time delay and the steady state gain, respectively.

\begin{small}
\begin{table*}[t]
 \centering
 \resizebox{\textwidth}{!}{
\begin{tabular}{|c|c|c|c|c|} \hline
Control systems & \# bits & \multicolumn{2}{c|}{Synthesized gains} &Time cost\\
\cline{3-4}
&&K&L&\\
\hline
Bicycle & 16 & [3.0253~12.6089] &$[0.0132~0.1021]^T$&1h36m41s\\
\hline
DC motor position  & 16 & [0.1129~0.0211~0.0093] &$[0.0390~0.3700~-0.0175]^T$& 1h39m06s\\
\hline
Pitch angle control & 32 & [ -0.1202~42.5655~1.0001] &$[0.0001~0~0.0017]^T$&8h31m53s\\
\hline
Inverted pendulum & 32 & [-1.5362~-2.0254~16.5192~2.7358]&$\left[ {\begin{array}{cccc} 0.0017&0.0021&0.0012&0\\0.0001&0.0018&0.0122&0.0770\end{array}}\right]^T$& 9h54m17s\\
\hline
Batch reactor process & 16 & $\left[ {\begin{array}{cccc} 0.0583&0.9093&0.3258&0.8721\\-2.4638&-0.0504&-1.7099&1.1653\end{array}}\right]$&$\left[ {\begin{array}{cccc} 0.0774&-0.0022&0.0267&0.0356 \\-0.0103&0.0227&0.0398&0.0001\end{array}}\right]^T$& 3h08m29s\\
\hline
\end{tabular}}
\caption{Synthesized gains and required time for synthesizing them.}
\label{table-exp1}
\end{table*}
\end{small}

\begin{small}
\begin{table*}[t]
\begin{center}
\resizebox{\textwidth}{!}{
 \begin{tabular}{|c | c | c | c | c | c | c |}\hline
Control& \multicolumn{2}{c|}{$lub$ of LQR cost}& \multicolumn{2}{c|}{LQG cost}& \multicolumn{2}{c|}{Steady state error} \\
\cline{2-7}
 systems& LQR & Synthesized & LQG & Synthesized & LQR-LQG & Synthesized \\
 &&K&&L&&gains\\
\hline
Bicycle&3956.3$\Vert{x}\Vert^2$&4331.7$\Vert{x}\Vert^2$&0.0229&0.0246&5.0489$b(e_1)$+0.5486&2.5341$b(e_1)$+0.0513\\
\hline
DC motor position&1001.6$\Vert{x}\Vert^2$&1376.7$\Vert{x}\Vert^2$&36.6315&36.6731&30.566$b(e_1)$+0.16&15.421$b(e_1)$+0.011\\
\hline
Pitch angle control&$2.9732\times10^6\Vert{x}\Vert^2$&$2.9887\times10^6\Vert{x}\Vert^2$&0.0013&0.0018&2.6781$b(e_1)$+0.4746&1.4453$b(e_1)$+0.0807\\
\hline
Inverted pendulum&$4.2988\times10^4\Vert{x}\Vert^2$&$5.3471\times10^4\Vert{x}\Vert^2$&0.3600&0.3897&83.4217$b(e_1)$+0.0432&30.3801$b(e_1)$+0.0086\\
\hline
Batch reactor process&223.1773$\Vert{x}\Vert^2$&223.1825$\Vert{x}\Vert^2$&0.0731&0.0949&2.9309$b(e_1)$+0.4194&2.1216$b(e_1)$+0.1642\\
\hline
\end{tabular}}
\caption{Least upper bound ($lub$) on the LQR cost (\ref{LQR_cost}), for a given initial condition $x$, the LQG cost (\ref{LQG}), and the Euclidean norm of the steady state error for the LQR-LQG and the synthesized gains.}
\label{table-exp2}
\end{center}
\end{table*}
\end{small}

\section{Experimental Results}
We implemented the algorithm presented in Section~\ref{algorithm} in Matlab. 
We use a PSO function in Matlab, introduced in~\cite{ebbesen}.
We implemented a static analyzer in Ocaml that synthesizes the best fixed-point program and computes the 
bound on the fixed-point implementation error for given feedback and observer gains $K$ and $L$, respectively. 
The tool gets the number of the bits in the fixed-point datatype, compact subsets $Y\subset\R^p$ and $\widehat{X}\subset\R^n$, 
and feedback and observer gains $K$ and $L$, respectively, as inputs. 
The optimization problems in computing the error bound are solved using the mixed-integer linear programming tool {\sf lp\_solve}~\cite{lpsolve}. 

We applied the proposed controller synthesis approach to a number of linear control systems. 
All the experiments were done on a laptop with CPU Intel Core 2 Duo at 2.4 GHz. 
In all of the experiments, the number of the particles in PSO is $N=24$, 
the maximum number of iterations is set to $l_{\max}=100$, 
and we choose the matrices $Q=I_n$, and $R=I_m$ in (\ref{LQR_cost}) and $\widehat{Q}=I_q$, and $\widehat{R}=I_p$ in (\ref{spectrum}). The value of $l_{\max}$ was chosen in such a way that appropriate gains are obtained in terms of the cost function (\ref{cost1}) (or (\ref{cost2})) for all control systems.
Moreover, we assume that the search space is $D=\prod_{i=1}^{n\times{m}+n\times{p}} [-150,~150]\subset\R^{n\times{m}+n\times{p}}$ that is large enough and contains the standard LQR and LQG gains for all the examples. Furthermore, without loss of generality, we work on the compact subsets $Y=\prod_{i=1}^p [-1,1]\subset\R^p$ and $\widehat{X}=\prod_{i=1}^n [-1,1]\subset\R^n$. All constants and variables are expressed in SI units. 

Our unstable examples include a bicycle~\cite{astrom}, a DC motor position control~\cite{cmu_examples}, a pitch angle control~\cite{cmu_examples}, an inverted pendulum~\cite{cmu_examples}, a batch reactor process~\cite{green} and another inverted pendulum for PID synthesis~\cite{cmu_examples}. See Table~\ref{table-exp1} and \ref{table-exp2} for experimental results. Note that for those examples for which 32-bit implementation is chosen, the 16-bit one provides a stability region which is even larger than the range of the variables inside the controller. As can be seen from Table \ref{table-exp2}, in comparison with the conventional LQR-LQG approach, the proposed synthesis approach in this paper worsens the LQR and LQG performances by at most $1.37$ times (for
DC motor position) and $1.38$ times (for Pitch angle control), respectively. 
However, the proposed synthesis approach improves the size of the region of practical stability due to quantization error by at least $2.55$ times. For certain examples, the improvement goes beyond the factor of 10. For bicycle  and DC motor position, the region of practical stability due to quantization error improves by a factor of $10.69$ and $14.55$, respectively. 

The detailed description of the systems are available as follows. 

\begin{figure}
\begin{center}
\includegraphics[width=4in]{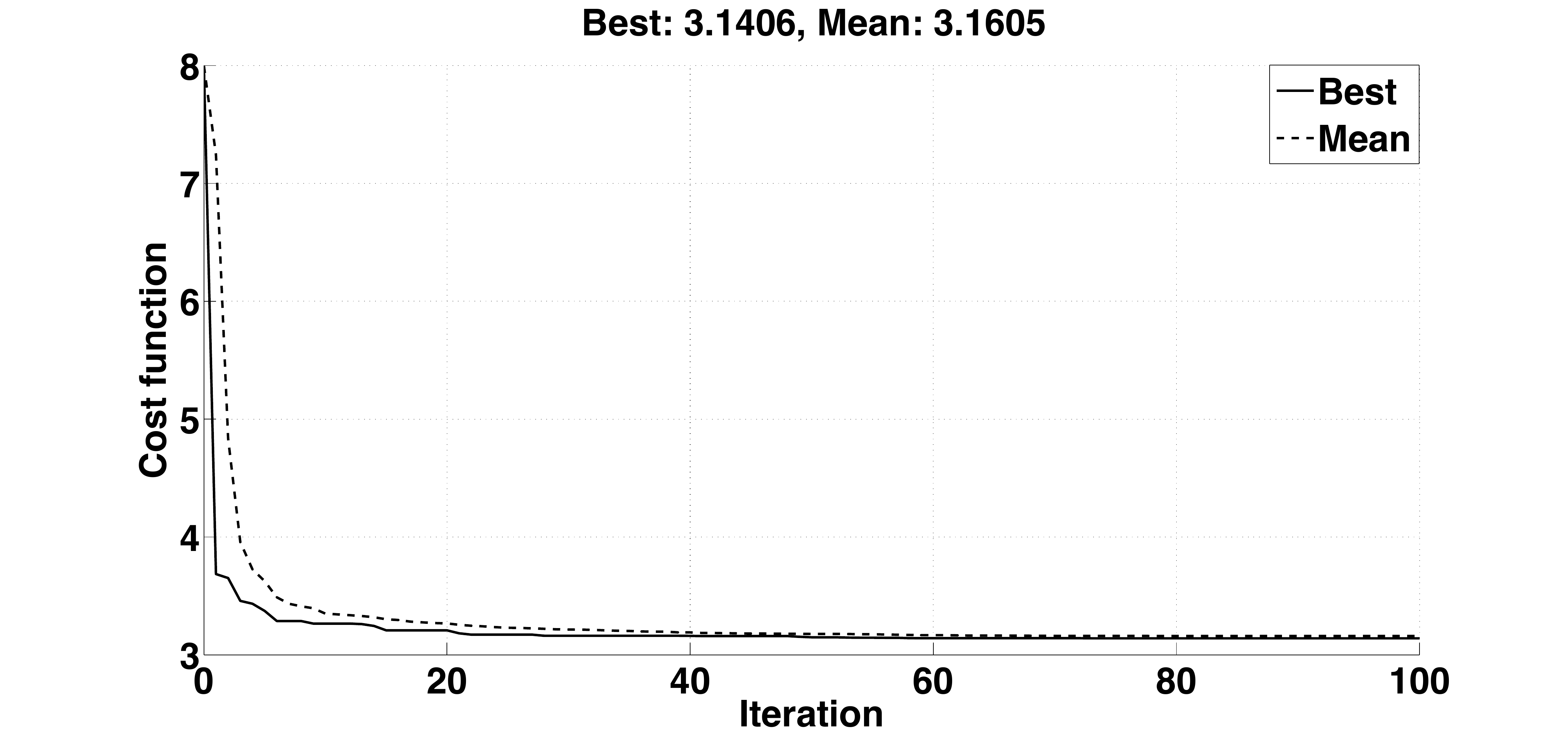}
\end{center}
\caption{Cost of the best particle and average cost of all population vs iteration.}
\label{fig2}
\end{figure}

\myparagraph{Bicycle}
The model of a bicycle is shown in (\ref{bicycle}). The weighting factors in (\ref{cost1}) are chosen as $w_1=w_2=w_3=1$ and $w_4=5$. The results of the LQR, LQG and the proposed method are shown in Tables \ref{table-exp1} and \ref{table-exp2}.
Figure~\ref{fig2} shows how the value of the cost function improves with the number of iteration. The figure shows how
the value of the cost function monotonically decreases with the number of iterations. The fixed-point C code
for the synthesized controller is shown in Figure~\ref{fig:fxc}.

\begin{figure}
\centering
\begin{tt}
\begin{scriptsize}
\parbox{0cm}{
\begin{tabbing}
floa\=t output(float yin)\ \ \=\\
\{\\
\>    static int $x1$ = $x1_0$;\>	// fixdt(1,16,14) \\
\>    static int $x2$ = $x2_0$;\>	// fixdt(1,16,14) \\
\>    int $x1\_new$;	\> 			// fixdt(1,16,14) \\
\>    int $x2\_new$;	\>			// fixdt(1,16,14) \\
\>    int $u$;		\>        		// fixdt(1,16,11) \\
\\    
\>    // Intermediate variables \\
\>    int $Gain1$;              \>		// fixdt(1,16,15) \\
\>    int $Gain2$;              \>		// fixdt(1,16,15) \\
\>    int $Gain3$;              \>		// fixdt(1,16,15) \\
\>    int $Add1$;               \>		// fixdt(1,16,14) \\
\>    int $Gain4$;              \>		// fixdt(1,16,15) \\
\>    int $Gain5$;              \>		// fixdt(1,16,15) \\
\>    int $Gain6$;              \>		// fixdt(1,16,15) \\
\>    int $Add2$;               \>		// fixdt(1,16,15) \\
\>    int $Gain7$;              \>		// fixdt(1,16,13) \\
\>    int $Gain8$;              \>		// fixdt(1,16,11) \\
\\    
\>    $y$ = convert\_to\_fixedpoint($yin$);  \\
\>    $Gain1$ = $(31499 * x1) >> 14$; \\
\>    $Gain2$ = $(-3145 * x2) >> 14$; \\
\>    $Add1$ = $(Gain1 + Gain2) >> 1$; \\
\>    $Gain3$ = $(432 * y) >> 14$; \\
\>    $x1\_new$ = $((Add1 << 1) + Gain3) >> 1$; \\
\>    $Gain4$ = $(-1907 * x1) >> 14$; \\
\>    $Gain5$ = $(23835 * x2) >> 14$; \\
\>    $Add2$ = $Gain4 + Gain5$; \\
\>    $Gain6$ = $(3345 * y) >> 14$; \\
\>    $x2\_new$ = $(Add1 + Gain6) >> 1$; \\
\>    $Gain7$ = $(24783 * x1\_new) >> 14$; \\
\>    $Gain8$ = $(25823 * x2\_new) >> 14$; \\
\>    $u$ = $(Gain7 + (Gain8 << 2)) >> 2$; \\
\>    return(float($u$)); \\ 
\}
\end{tabbing}}
\end{scriptsize}
\end{tt}
\caption{synthesized fixed-point controller C code for Bicycle.} \label{fig:fxc}
\end{figure}

\myparagraph{DC motor position control}
The dynamic of a DC motor position control, borrowed from \cite{cmu_examples}, is given by:
\begin{align}
\nonumber
\left\{\begin{array}{l}\left[{\begin{array}{c}
\dot\xi_1\\
\dot\xi_2\\
\dot\xi_3\\
 \end{array}}\right]=\left[ {\begin{array}{ccc}
0&1&0\\
0&\frac{-b}{J}&\frac{K}{J}\\
0&\frac{-K}{L}&\frac{-R}{L}\\
 \end{array}}\right]\left[{\begin{array}{c}
\xi_1\\
\xi_2\\
\xi_3\\
 \end{array}}\right]+\left[ {\begin{array}{c}
0\\
0\\
\frac{1}{L}\\
 \end{array}}\right]\left(\upsilon+\omega\right),\\\hspace{10mm}\eta=[1~~0~~0]\left[{\begin{array}{c}
\xi_1\\
\xi_2\\
\xi_3\\
 \end{array}}\right]+\nu,\end{array}\right. 
\end{align}
where $\xi_1$ is the angle of the motor's shaft, $\xi_2$ is the angular velocity of the motor's shaft, $\xi_3$ is the armature current, $b=3.508\times{10^{-6}}$ is the damping ratio of the mechanical system, $J=3.228\times{10^{-6}}$ is the moment of inertia of the rotor, $K=0.027$ is the electromotive force constant, $R=4$ is the electric resistance, $L=2.75\times{10^{-6}}$ is the electric inductance and $\upsilon$ is the source voltage. The weighting factors in (\ref{cost1}) are chosen as $w_1=w_2=w_3=1$ and $w_4=5$. The LQR and LQG gains are given by \mbox{$K_{LQR}=[0.4055~~0.3782~~0.0022]$} and \mbox{$L_{LQG}=[0.0288~~0.3858~~-0.0026]^T$} and the gains, computed by the proposed approach in this paper, are given in Table \ref{table-exp1}. The detailed results are shown in Tables \ref{table-exp1} and \ref{table-exp2}. 

\myparagraph{Pitch control}
The dynamic of the longitudinal motion of an aircraft, borrowed from \cite{cmu_examples}, is given by:
\begin{align}
\nonumber
\left\{\begin{array}{l}\left[{\begin{array}{c}
\dot\xi_1\\
\dot\xi_2\\
\dot\xi_3\\
 \end{array}}\right]=\left[ {\begin{array}{ccc}
-0.313&56.7&0\\
-0.0139&-0.426&0\\
0&56.7&0\\
 \end{array}}\right]\left[{\begin{array}{c}
\xi_1\\
\xi_2\\
\xi_3\\
 \end{array}}\right]+\left[ {\begin{array}{c}
0.232\\ 0.0203\\ 0\\
 \end{array}}\right]\left(\upsilon+\omega\right)\\\hspace{10mm}\eta=[0~~0~~1]\left[{\begin{array}{c}
\xi_1\\
\xi_2\\
\xi_3\\
 \end{array}}\right]+\nu,\end{array}\right. 
\end{align}
where $\xi_1$ is the angle of attack, $\xi_2$ is the pitch rate, $\xi_3$ is the pitch angle, and $\upsilon$ is elevator deflection angle. The weighting factors in (\ref{cost1}) are chosen as $w_1=w_2=w_3=1$ and $w_4=5$. The LQR and LQG gains are given by $K_{LQR}=[-0.1141~~49.1428~~0.9995]$ and $L_{LQG}=10^{-3}\times[0.6407~~0.0039~~0.6655]^T$ and the gains, computed by the proposed approach in this paper, are given in Table \ref{table-exp1}. The detailed results are shown in Tables \ref{table-exp1} and \ref{table-exp2}. 

\myparagraph{Inverted pendulum}
Consider a simple physical model of an inverted pendulum on a cart, borrowed from \cite{cmu_examples}. The dynamics of the system is given by:
\begin{align}
\nonumber
\left\{\begin{array}{l}\left[{\begin{array}{c}
\dot\xi_1\\
\dot\xi_2\\
\dot\xi_3\\
\dot\xi_4\\
 \end{array}}\right]=\left[ {\begin{array}{cccc}
0&1&0&0 \\
0&\frac{-\left(I+ml^2\right)b}{I(M+m)+Mml^2}&\frac{m^2gl^2}{I(M+m)+Mml^2}&0\\
0&0&0&1\\
0&\frac{-mlb}{I(M+m)+Mml^2}&\frac{mgl(M+m)}{I(M+m)+Mml^2}&0
 \end{array}}\right]\left[{\begin{array}{c}
\xi_1\\
\xi_2\\
\xi_3\\
\xi_4
 \end{array}}\right]+\left[ {\begin{array}{c}
0\\
\frac{I+ml^2}{I(M+m)+Mml^2}\\
0\\
\frac{ml}{I(M+m)+Mml^2}\\
 \end{array}}\right]\upsilon+\left[ {\begin{array}{c}
1\\
1\\
1\\
1\\
 \end{array}}\right]\omega,\\\hspace{10mm}\eta=\left[{\begin{array}{cccc}
1&0&0&0\\
0&0&1&0\\
 \end{array}}\right]\left[{\begin{array}{c}
\xi_1\\
\xi_2\\
\xi_3\\
\xi_4
 \end{array}}\right]+\nu,\end{array}\right. 
\end{align}
where $\xi_1$, and $\xi_2$ are the position and velocity of the cart, respectively, $\xi_3$, and $\xi_4$ are the angular position and velocity of the mass to be balanced, $\upsilon$ is the applied force to the cart, $g=9.8$ is the acceleration due to gravity, $l=0.3$ is the length of the rod, $m=0.2$ is the mass of the system to be balanced, $M=0.5$ is the mass of the cart, $b=0.1$ is the coefficient of friction of the cart, and $I=0.006$ is the inertia of the pendulum. The weighting factors in (\ref{cost1}) are chosen as $w_1=w_2=w_3=1$ and $w_4=5$. The LQR and LQG gains are given by $K_{LQR}=[-0.9929~~-2.0276~~20.2819~~3.9126]$ and 
\begin{align}\nonumber
L_{LQG}=\left[{\begin{array}{cccc}
0.0016&0.0011&0.0007&0.0034\\
0.0007&0.0051&0.0111&0.0618\\
 \end{array}}\right]^T,
\end{align}
and the gains, computed by the proposed approach in this paper, are given in Table \ref{table-exp1}. The detailed results are shown in Tables \ref{table-exp1} and \ref{table-exp2}. 

\myparagraph{Batch reactor process}
Consider an unstable batch reactor process, borrowed from \cite{green}. The dynamic of the system is given by:
\begin{align}
\nonumber
\left\{\begin{array}{l}\left[{\begin{array}{c}
\dot\xi_1\\
\dot\xi_2\\
\dot\xi_3\\
\dot\xi_4\\
 \end{array}}\right]=\left[ {\begin{array}{cccc}
1.38&-0.2077&6.715&-5.676 \\
-0.5814&-4.29&0&0.675\\
1.067&4.273&-6.654&5.893\\
0.048&4.273&1.343&-2.104\\
 \end{array}}\right]\left[{\begin{array}{c}
\xi_1\\
\xi_2\\
\xi_3\\
\xi_4
 \end{array}}\right]+\left[ {\begin{array}{cc}
0&0\\
5.679&0\\
1.136&-3.146\\
1.136&0\\
 \end{array}}\right]\upsilon+\left[ {\begin{array}{c}
1\\
1\\
1\\
1\\
 \end{array}}\right]\omega,\\
 \hspace{10mm}\eta=\left[ {\begin{array}{cccc}
1&0&1&-1\\
0&1&0&0\\
 \end{array}}\right]\left[{\begin{array}{c}
\xi_1\\
\xi_2\\
\xi_3\\
\xi_4
 \end{array}}\right]+\nu.\end{array}\right.
\end{align}
The weighting factors in (\ref{cost1}) are chosen as $w_1=w_3=1$, $w_2=2$, and $w_4=5$. The LQR and LQG gains are given by:  
\begin{align}\nonumber
K_{LQR}=&\left[{\begin{array}{cccc}
0.0376&0.9157&0.3262&0.8226\\
-2.4884&-0.0734&-1.7461&1.1438\\
 \end{array}}\right],\\\notag
L_{LQG}=&\left[{\begin{array}{cccc}
0.0447&-0.0003&0.0170&0.0127\\
0&0.0020&0.0058&0.0059\\
 \end{array}}\right]^T,
\end{align}
and the gains, computed by the proposed approach in this paper, are given in Table \ref{table-exp1}. The detailed results are shown in Tables \ref{table-exp1} and \ref{table-exp2}. 

\myparagraph{PID controller}
In this example, we provide a PID controller for an inverted pendulum whose dynamic is given by a transfer function. Consider the transfer function of an inverted pendulum, borrowed from \cite{cmu_examples}, given by:
\begin{equation}
\frac{\Phi(s)}{U(s)}={\frac{\frac{ml}{q}s}{s^3+\frac{b(I+ml^2)}{q}s^2-\frac{(M+m)mgl}{q}s-\frac{bmgl}{q}}},
\end{equation}
where $q=(M+m)(I+ml^2)-(ml)^2$, output $\phi$ is the angular position of the mass to be balanced, input $\upsilon$ is the applied force to the cart, $g=9.8$ is the acceleration due to gravity, $l=0.3$ is the length of the rod, $m=0.2$ is the mass of the system to be balanced, $M=0.5$ is the mass of the cart, $b=0.1$ is the coefficient of friction of the cart, and $I=0.006$ is the inertia of the pendulum. Using standard results in control theory \cite{kailath}, one obtains the following state space realization for the inverted pendulum:
\begin{align}
\nonumber
\left\{\begin{array}{l}\left[{\begin{array}{c}
\dot\xi_1\\
\dot\xi_2\\
\dot\xi_3\\
 \end{array}}\right]=\left[ {\begin{array}{ccc}
-0.1818&3.8977&0.5568\\
8.000&0&0\\
0&1&0\\
 \end{array}}\right]\left[{\begin{array}{c}
\xi_1\\
\xi_2\\
\xi_3\\
 \end{array}}\right]+\left[ {\begin{array}{c}
1\\0\\ 0\\
 \end{array}}\right]\upsilon,\\\hspace{10mm}\phi=[0~~0.5682~~1]\left[{\begin{array}{c}
\xi_1\\
\xi_2\\
\xi_3\\
 \end{array}}\right].\end{array}\right. 
\end{align}  

Our objective is to design PID gains $K_P$, $K_I$, and $K_D$ minimizing the cost function (\ref{cost2}) with weighting factors $w_1=w_2=w_3=1$ and the closed loop system has a settling time ($t_s$) of less than 5 seconds and pendulum should not move more than 0.05 radians away from the vertical axis. The latter two constrains are treated the same as the stability constraint in Subsection \ref{algorithm} by penalizing the cost function (\ref{cost2}). The synthesized gains are $K_P=109.032$, $K_I=1.2268$, and $K_D=13.9945$. The closed loop system has $PM=+\infty$, $GM=26237$, $\gamma(b(e_{q1})+b(e_{q2}))=4.1705\times10^{-4}$, settling time $t_s=0.4790$, and pendulum does not move more than 0.0098 radians away from the vertical axis.

\section{Conclusion}

We have presented a generic methodology to search for optimal controller implementations that
minimize implementation errors in addition to traditional controller performance criteria.
While we have instantiated the methodology using the LQR and LQG costs and quantization errors, our algorithm
is more generally applicable to other performance criteria and other sources of modeling or implementation error.
The controller synthesis algorithm can be seamlessly added to any design and automatic code generation tool flow
to enhance its capability to generate correct-by-construction high performance controller software. 
By automatically synthesizing the minimal error controller, we sidestep the need for post-design verification.

\bibliographystyle{alpha}
\bibliography{reference,main}

\newcommand{\etalchar}[1]{$^{#1}$}
\begin{thebibliography}{ZKGSP09}

\bibitem[AM08]{astrom}
K.~J. Astrom and R.~M. Murray.
\newblock {\em Feedback systems}.
\newblock Princeton University Press, 2008.

\bibitem[AMST10]{anta}
A.~Anta, R.~Majumdar, I.~Saha, and P.~Tabuada.
\newblock Automatic verification of control system implementations.
\newblock {\em In proceedings of EMSOFT}, pages 9--18, October 2010.

\bibitem[BR05]{BR_RSP_05}
P.~Belanovic and M.~Rupp.
\newblock {A}utomated {F}loating-point to {F}ixed-point {C}onversion with the
  {F}ixify {E}nvironment.
\newblock In {\em Proc. International Workshop on Rapid System Prototyping},
  pages 172--178, 2005.

\bibitem[CF95]{chen}
T.~Chen and B.~A. Francis.
\newblock {\em Optimal sampled-data control systems}.
\newblock Springer-Verlag, New York, 1995.

\bibitem[cmu]{cmu_examples}
Control tutorial for matlab and simulink.
\newblock {\em Available online at http://www.library.cmu.edu/ctms/ctms/}.

\bibitem[DM11]{DM11}
P.~S. Duggirala and S.~Mitra.
\newblock Abstraction-refinement for stability.
\newblock {\em in Proceedings of ICCPS}, pages 22--31, April 2011.

\bibitem[EKG12]{ebbesen}
S.~Ebbesen, P/~Kiwitz, and L.~Guzzella.
\newblock A generic particle swarm optimization function for {M}atlab.
\newblock {\em American Control Conference (to appear)}, June 2012.

\bibitem[Fer10]{feron_journal}
Eric Feron.
\newblock From control systems to control software.
\newblock {\em IEEE Control Systems Magazine}, 30(6):50--71, December 2010.

\bibitem[GL94]{green}
M.~Green and D.~J.~N. Limebeer.
\newblock {\em Linear robust control}.
\newblock Prentice Hall, August 1994.

\bibitem[Hes09]{joao}
J.~P. Hespanha.
\newblock {\em Linear systems theory}.
\newblock Princeton University Press, September 2009.

\bibitem[JLY07]{jiang}
M.~Jiang, Y.~P. Luo, and S.~Y. Yang.
\newblock Stochastic convergence analysis and parameter selection of the
  standard particle swarm optimization algorithm.
\newblock {\em Information Processing Letters}, 102(1):8--16, April 2007.

\bibitem[Kai80]{kailath}
T.~Kailath.
\newblock {\em Linear systems}.
\newblock Prentice-Hall, Inc., 1980.

\bibitem[Kal56]{kalman}
R.~E. Kalman.
\newblock Nonlinear aspects of sampled-data control systems.
\newblock {\em in Proceedings of the Symposium on Nonlinear Circuit Analysis,
  edited by J. Fox, Polytechnic Institute of Brooklyn}, pages 273--313, 1956.

\bibitem[KE95]{kennedy}
J.~Kennedy and R.~Eberhart.
\newblock Particle swarm optimization.
\newblock {\em In Proceedings of IEEE International Conference on Neural
  Networks}, pages 1942--1948, 1995.

\bibitem[LAS09]{liu}
H.~Liu, A.~Abraham, and V.~Snasel.
\newblock Convergence analysis of swarm algorithm.
\newblock {\em World congress on Nature and Biologically Inspired Computing},
  pages 1714--1719, December 2009.

\bibitem[LCNT07]{LCN_TCAD_07}
J.~A. L$\acute{o}$pez, C.~Carreras, and O.~Nieto-Taladriz.
\newblock {I}mproved {I}nterval-based {C}haracterization of {F}ixed-point {LTI}
  {S}ystems with {F}eedback {L}oops.
\newblock {\em IEEE Trans. on CAD of Integrated Circuits and Systems},
  26(11):1923--1932, 2007.

\bibitem[LGC{\etalchar{+}}06]{LGCMLC_TCAD_06}
D.~Lee, A.~A. Gaffar, R.~C.~C. Cheung, O.~Mencer, W.~Luk, and G.~A.
  Constantinides.
\newblock {A}ccuracy-{G}uaranteed {B}it-width {O}ptimization.
\newblock {\em IEEE Trans. on CAD of Integrated Circuits and Systems},
  25(10):1990--2000, 2006.

\bibitem[lps]{lpsolve}
lp\_solve, a {M}ixed {I}nteger {L}inear {P}rogramming ({MILP}) solver.
\newblock {\em Available online at http://lpsolve.sourceforge.net/}.

\bibitem[LSG92]{liu1}
K.~Liu, R.~E. Skelton, and K.~Grigoriadis.
\newblock Optimal controllers for finite wordlength implementation.
\newblock {\em IEEE Transactions on Automatic Control}, 37(9):1294--1304,
  September 1992.

\bibitem[LSW96]{lin}
Y.~Lin, E.~D. Sontag, and Y.~Wang.
\newblock A smooth converse lyapunov theorem for robust stability.
\newblock {\em SIAM Journal on Control and Optimization}, 34(1):124--160, 1996.

\bibitem[Moo66]{Moore66}
R.~Moore.
\newblock {\em {I}nterval {A}nalysis}.
\newblock Prentice Hall, 1966.

\bibitem[MSBZ07]{MSBZ_TCAD_07}
A.~Mallik, D.~Sinha, P.~Banerjee, and H.~Zhou.
\newblock {L}ow-{P}ower {O}ptimization by {S}mart {B}it-width {A}llocation in a
  {S}ystem{C}-based {ASIC} {D}esign {E}nvironment.
\newblock {\em IEEE Trans. on CAD of Integrated Circuits and Systems},
  26(3):447--455, 2007.

\bibitem[OCC{\etalchar{+}}07]{OCCLM_FPL_07}
W.~G. Osborne, R.~C.~C. Cheung, J.~G.~F. Coutinho, W.~Luk, and O.~Mencer.
\newblock {A}utomatic {A}ccuracy-{G}uaranteed {B}it-width {O}ptimization for
  {F}ixed and {F}loating-point {S}ystems.
\newblock In {\em Proc. FPL}, pages 617--620, 2007.

\bibitem[PW06]{PodelskiW06}
A.~Podelski and S.~Wagner.
\newblock Model checking of hybrid systems: from reachability towards
  stability.
\newblock {\em in Proceedings of HSCC}, pages 507--521, April 2006.

\bibitem[PW07]{PodelskiW07}
Podelski and Wagner.
\newblock Region stability proofs for hybrid systems.
\newblock {\em in Proceedings of FORMATS}, pages 320--335, 2007.

\bibitem[SF97]{SF_IMPA_97}
J.~Stolfi and L.~H. Figueiredo.
\newblock {S}elf-validated {N}umerical {M}ethods and {A}pplications.
\newblock In {\em Monograph for 21st Brazilian Mathematics Colloquium, Rio de
  Janeiro: IMPA}, 1997.

\bibitem[Wil85]{williamson}
D.~Williamson.
\newblock Finite wordlength design of digital {K}alman filters for state
  estimation.
\newblock {\em IEEE Transactions on Automatic Control}, 30(10):930--939,
  October 1985.

\bibitem[Wil89]{williamson1}
D.~Williamson.
\newblock Optimal finite wordlength linear quadratic regulation.
\newblock {\em IEEE Transactions on Automatic Control}, 34(12):1218--1228,
  December 1989.

\bibitem[Win93]{Winskel}
G.~Winskel.
\newblock {\em The formal semantics of programming languages: an introduction}.
\newblock MIT Press, February 1993.

\bibitem[ZKGS07]{majid2}
M.~Zamani, M.~Karimi-Ghartemani, and N.~Sadati.
\newblock {FOPID} controller design for robust performance using particle swarm
  optimization.
\newblock {\em Journal of Fractional Calculus $\&$ Applied Analysis (FCAA)},
  10(2):169--188, 2007.

\bibitem[ZKGSP09]{majid}
M.~Zamani, M.~Karimi-Ghartemani, N.~Sadati, and M.~Parniani.
\newblock Design of a fractional order {PID} controller for an {AVR} using
  particle swarm optimization.
\newblock {\em Control Engineering Practice}, 17(12):1380--1387, December 2009.

\bibitem[ZSKG09]{majid1}
M.~Zamani, N.~Sadati, and M.~Karimi-Ghartemani.
\newblock Design of an ${H}_\infty$ {PID} controller using particle swarm
  optimization.
\newblock {\em International Journal of Control, Automation, and Systems
  (IJCAS)}, 7(2):273--280, April 2009.

\end{thebibliography}

\end{document}